\documentclass[11pt]{article}
\usepackage[utf8]{inputenc}
\usepackage[margin=1in]{geometry}
\usepackage[authoryear, round]{natbib}
\usepackage{graphicx}
\usepackage{amsthm}
\usepackage{amsmath}
\usepackage{amssymb}
\usepackage{placeins}
\usepackage{textcomp,gensymb}
\usepackage{caption, subcaption}
\usepackage{comment}
\usepackage{booktabs}
\usepackage{adjustbox}
\usepackage{color}
\usepackage{url}
\usepackage{tikz}
\usetikzlibrary{bayesnet}

\usepackage{setspace}
\usepackage{titlesec}

\titleformat*{\section}{\large\bfseries}
\titleformat*{\subsection}{\bfseries}
\titleformat*{\subsubsection}{\bfseries}
\titleformat*{\paragraph}{\itshape}
\titleformat*{\subparagraph}{\bfseries}

\title{The Problem of Infra-marginality\\in Outcome Tests for Discrimination\thanks{We 
thank Cheryl Phillips and Vignesh Ramachandran of the Stanford Computational Journalism Lab for helping to compile the North Carolina traffic stop data,
and the John S. and James L. Knight Foundation for partial support of this research.
We also thank Stefano Ermon, Avi Feller, Seth Flaxman, Andrew Gelman, Lester Mackey, Jan Overgoor, and Emma Pierson for helpful comments.
Our dataset of North Carolina traffic stops is available at \protect\url{https://purl.stanford.edu/nv728wy0570},
and our analysis code is available at \protect\url{https://github.com/5harad/threshold-test}.
}}
\author{Camelia Simoiu\\Stanford University \and Sam Corbett-Davies\\Stanford University \and Sharad Goel\\Stanford University}
\date{}

\begin{document}

\singlespacing
\maketitle
\thispagestyle{empty}

\begin{abstract}
Outcome tests are a popular method for detecting bias in lending, hiring, and policing decisions.
These tests operate by comparing the success rate of decisions across groups.
For example, if loans made to minority applicants are observed to be repaid more often than loans made
to whites, it suggests that only exceptionally qualified minorities are granted loans,
indicating discrimination.
Outcome tests, however, are known to suffer from the 
problem of \emph{infra-marginality}:
even absent discrimination,
the repayment rates for minority and white loan recipients might
differ if the two groups have different risk distributions.
Thus, at least in theory, outcome tests can fail to accurately detect discrimination.
We develop a new statistical test of discrimination---the threshold test---that mitigates 
the problem of infra-marginality by jointly estimating decision thresholds and risk distributions.
Applying our test to a dataset of 4.5 million police stops in North Carolina,
we find that the problem of infra-marginality is more than a theoretical possibility,
and can cause the outcome test to yield misleading results in practice.
\end{abstract}

\newpage
\setcounter{page}{1}

\section{Introduction}

Claims of biased decision making are typically hard to rigorously assess, 
in large part because of well-known problems with the two most common statistical tests for discrimination.
In the first test, termed \emph{benchmarking}, one compares the rate at which whites and minorities are
treated favorably. 
For example, in the case of lending decisions, if white applicants are granted loans more often
than minority applicants, that may be the result of bias against minorities.
However, if minorities in reality are less creditworthy than whites,
then such disparities in lending rates may simply reflect reasonable business practices rather
than discrimination.
This limitation of benchmarking is referred to in the literature as the \emph{qualified pool} or \emph{denominator} problem~\citep{ayres2002},
and is a specific instance of omitted variable bias.

Ideally, one would like to compare similarly qualified white and minority applicants,
but such a comparison requires detailed individual-level data and is often infeasible to carry out in practice.
Addressing this shortcoming of benchmarking, \citet{becker1993,becker1957} proposed the \emph{outcome test}, which is based not on the rate at which decisions are made,
but on the success rate of those decisions. 
Becker argued that even if minorities are less creditworthy than whites, minorities who are granted loans,
absent discrimination, should still be found to repay their loans at the same rate as whites who are granted loans.
If loans to minorities have a higher repayment rate than loans to whites, it suggests that lenders are applying a double standard,
granting loans only to exceptionally qualified minorities.
Though originally proposed in the context of lending decisions, 
outcome tests have gained popularity in a variety of domains,
particularly policing~\citep{goel2016,goel2016b,ayres2002,knowles2001}.
For example, when assessing bias in traffic stops,
one can compare the rates at which searches of white and minority drivers turn up contraband.
If searches of minorities yield contraband less often than searches of whites,
it suggests that the bar for searching minorities is lower, indicative of discrimination.

Outcome tests, however, are imperfect barometers of bias.
To see this, suppose that there are two, easily distinguishable types of white drivers: 
those who have  a 1\% chance of carrying contraband, 
and those who have a 75\% chance.
Similarly assume that black drivers have either a 1\% or 50\% chance of carrying contraband.
If officers, in a race-neutral manner, search individuals who are at least 10\% likely to be carrying contraband, 
then searches of whites will be successful 75\% of the time whereas searches of blacks will be successful only 50\% of the time.
This simple example illustrates a subtle failure of outcome tests
known as the problem of \emph{infra-marginality}~\citep{ayres2002}, a phenomenon we 
discuss in detail below.

Our contribution in this paper is two-fold.
First, we develop a new test for discrimination---the \emph{threshold test}---that mitigates
theoretical limitations of both benchmark and outcome analysis.
Our test simultaneously estimates decision thresholds and risk distributions
by fitting a hierarchical Bayesian latent variable model~\citep{gelman2014}.
In developing this method, we clarify the statistical origins of the problem of infra-marginality.
Second, we demonstrate that infra-marginality is more than a theoretical possibility, and
can cause the outcome test to yield misleading results in practice.
To do so, we analyze police vehicle searches 
in a dataset of 4.5 million traffic stops conducted by the 100 largest police departments in North Carolina.

\paragraph{Related work.} 

As the statistical literature on discrimination is extensive, we focus our review on policing.
Benchmark analysis is the most common statistical method for assessing racial bias in police stops and searches.
The key methodological challenge with this approach is estimating the race distribution of the at-risk, or benchmark, population.
Traditional benchmarks include the residential population, licensed drivers, arrestees, and reported crime suspects~\citep{engel2004}.
\citet{alpert2004}~estimate the race distribution of drivers on the roadway by considering not-at-fault drivers involved in two-vehicle crashes.
Others have looked at stops initiated by 
aerial patrols~\citep{mcconnell2001}, and those based on radar and cameras~\citep{lange2001}, arguing that such stops are less prone to potential bias and thus more likely to reflect the true population of traffic violators.
Studying police stops of pedestrians in New York City, \citet{gelman2007} use a hierarchical Bayesian model to construct a benchmark based on neighborhood- and race-specific crime rates. 
\citet{ridgeway2006} studies post-stop police actions by creating benchmarks based on propensity scores,
with minority and white drivers matched using demographics and the time, location, and purpose of the stops.
\citet{grogger2006} construct benchmarks
by considering stops at night, when a ``veil of darkness'' masks race.
 \citet{antonovics2009} use officer-level demographics in a variation of the standard benchmark test:
they argue that search rates that are higher when the officer's race differs from that of the suspect is evidence of discrimination.
Finally, ``internal benchmarks'' have been used to flag potentially biased officers by comparing each officer's stop decisions
to those made by others patrolling the same area at the same time~\citep{ridgeway2009,walker2003}.

Given the inherent limitations of benchmark analysis,
researchers have more recently turned to outcome tests to investigate claims of police discrimination.
For example, \citet{goel2016} use outcome analysis to test for racial bias in New York City's stop-and-frisk policy.
While outcome tests mitigate the problem of omitted variables faced by benchmark analysis,
they suffer from their own limitations, most notably infra-marginality.
The problem of infra-marginality in outcome tests was first discussed in detail by \citet{ayres2002}, 
although previous studies of discrimination~\citep{galster1993,carr1993} indicate awareness of the issue.  
An early attempt to address the problem was presented by \citet{knowles2001},
who developed an economic model of behavior in which drivers balance their utility for carrying contraband with the risk of getting caught, while officers balance the utility of finding contraband with the cost of searching.
Under equilibrium behavior, \citeauthor{knowles2001} argue that 
the hit rate (i.e., the search success rate) is identical to the search threshold, and so 
one can reliably detect discrimination with the standard outcome test.
\citet{engel2008} note that the model of  \citeauthor{knowles2001} requires strong assumptions, 
including that drivers and officers are rational actors, and that every driver has perfect knowledge of the likelihood that he will be searched.
\citet{anwar2006} propose a hybrid test of discrimination that is based on the rankings 
of race-contingent search and hit rates as a function of officer race: 
if officers are not prejudiced, they argue, then these rankings should be independent of officer race.
This approach circumvents the problems of omitted variables and infra-marginality in certain cases,
but it cannot detect discrimination when officers of different races are similarly biased.

\section{A New Test for Discrimination}
\label{sec:model}

\subsection{A model of decision making}
\label{sec:behavior}

We begin by introducing a stylized model of decision making that is the basis of our statistical approach,
and which also illustrates the problem of infra-marginality.
We develop this framework in the context of police stops, though the model itself applies more generally.

During routine traffic stops, officers have latitude to search both driver and vehicle for drugs, weapons, and other contraband
when they suspect more serious criminal activity.
These decisions are based on a myriad of contextual factors visible to officers during stops, 
including a driver's age and gender, criminal record, 
and behavioral indicators of nervousness of evasiveness.
We assume that officers use this information to estimate the probability a driver is carrying contraband,
and then conduct a search when that probability exceeds a fixed, race-specific search threshold
$t_r$.
Under this model, 
if officers have a lower threshold for searching blacks than whites (i.e., $t_{\text{black}} < t_{\text{white}}$),
then we would say that black drivers are being discriminated against.
Conversely, if $t_{\text{white}} < t_{\text{black}}$, we would say that white drivers are being discriminated against.
And if the thresholds are approximately equal across race groups, we would say there is no discrimination in search decisions.
In the economics literature, this is often referred to as \emph{taste-based discrimination}~\citep{becker1957}.\footnote{Taste-based discrimination
stands in contrast to \emph{statistical discrimination}~\citep{arrow1973,phelps1972}, in which officers might use a driver's race to improve their 
estimate that he is carrying contraband. Regardless of whether such information increases the efficiency of searches, 
officers are legally barred from using race to inform search decisions outside of circumscribed situations 
(e.g., when acting on specific and reliable suspect descriptions that include race among other factors).
As is standard in the empirical literature on racial bias, we test only for taste-based discrimination.}
We treat both the probabilities and the search thresholds as latent, unobserved quantities, and our goal is to 
infer them from data.

\begin{figure}[t!]
     \begin{subfigure}[t]{.48\linewidth}
         \centering
         \includegraphics[height=3in]{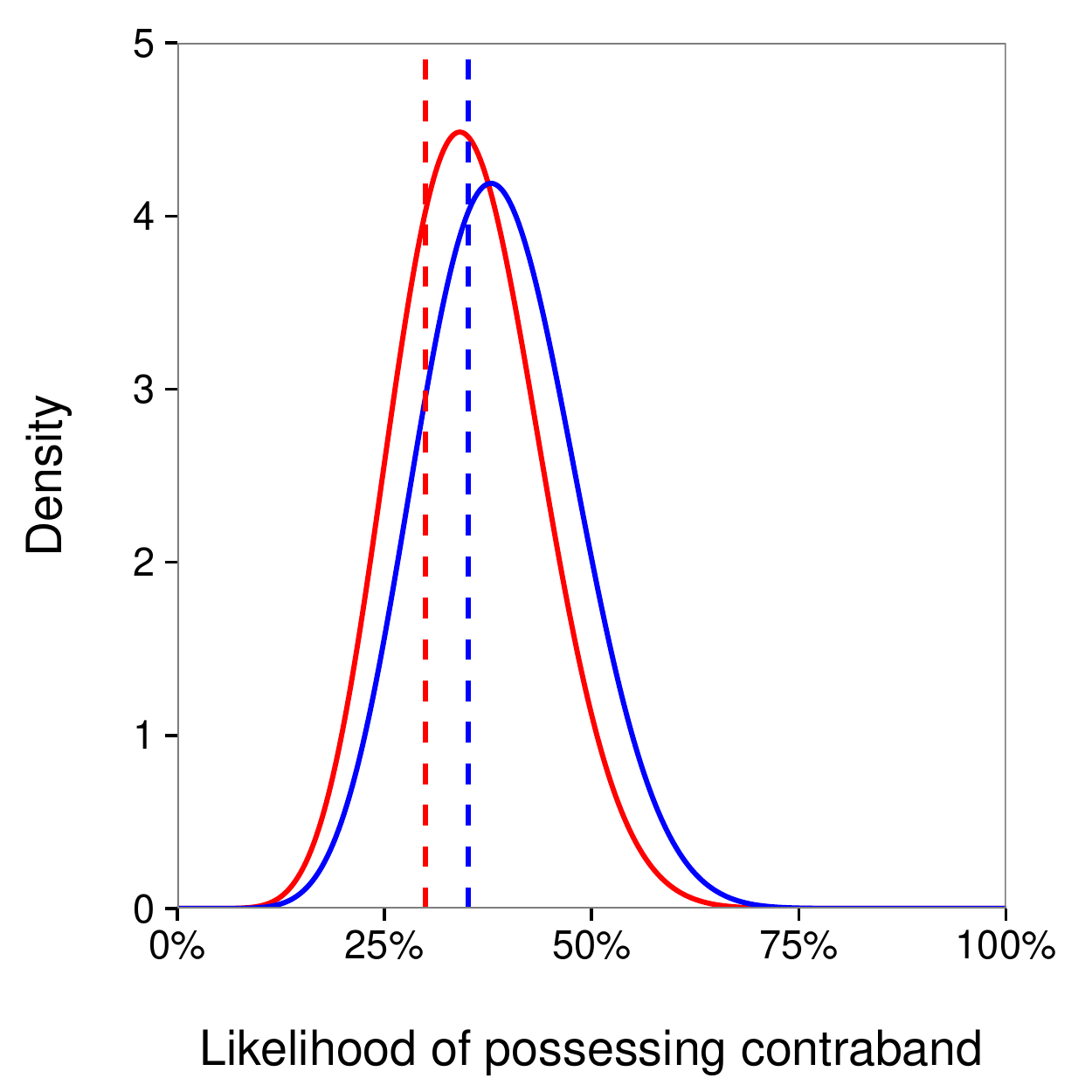}
         \caption{}
     \end{subfigure}
          \hspace{0.04\linewidth}
     \begin{subfigure}[t]{.48\linewidth}
 		\centering
 		\includegraphics[height=3in]{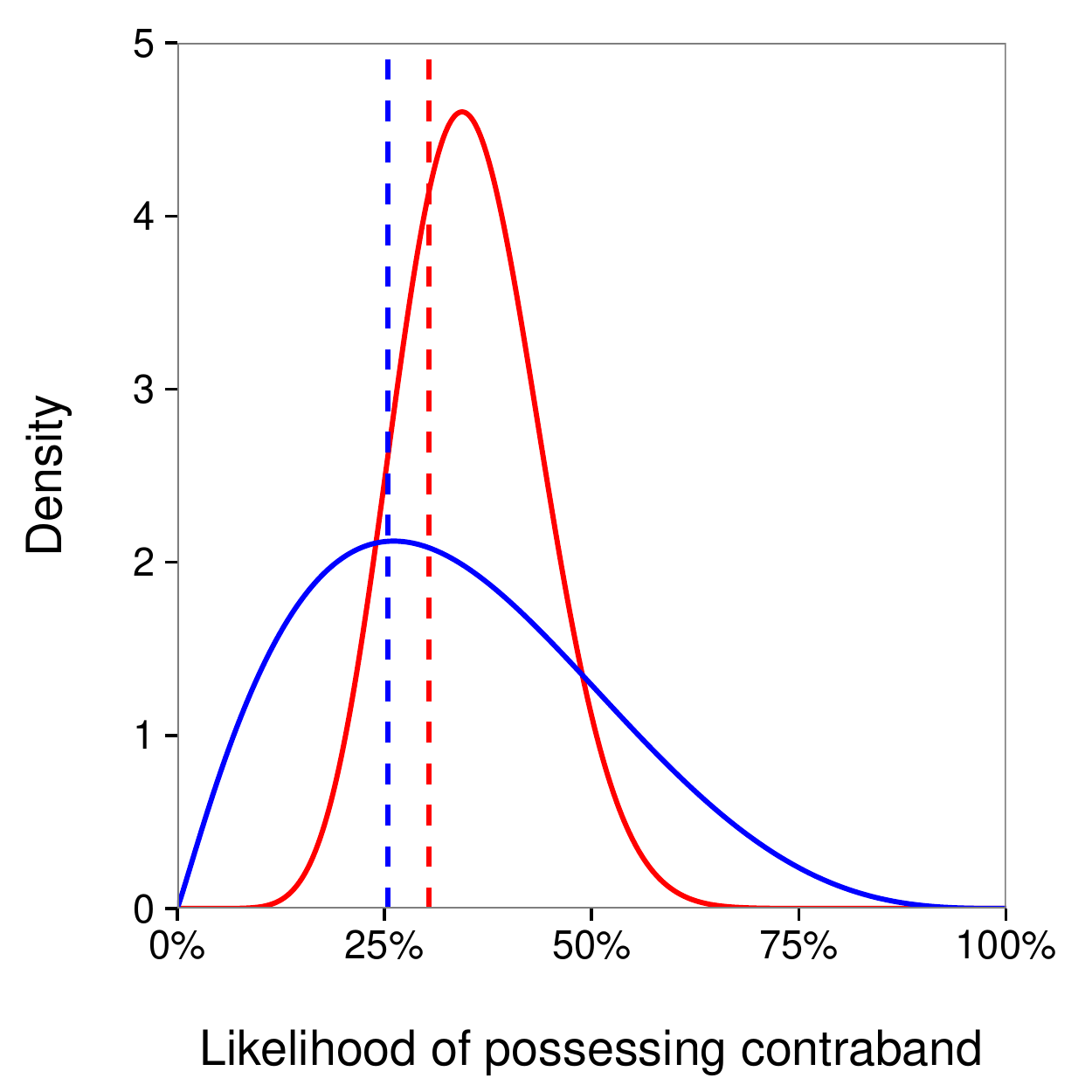}
 		\caption{}
     \end{subfigure}
     \caption[Example]{\emph{Hypothetical signal distributions
(solid curves) and search thresholds (dashed vertical lines) that illustrate how the benchmark and outcome tests can give misleading results.\footnotemark Under the model of Section~\ref{sec:behavior}, the search rate for a given group is equal to the area under the signal distribution above the threshold, and the hit rate is the mean of the distribution conditional on being above the  threshold.
Situations (a) and (b) are observationally equivalent: 
in both cases, red drivers are searched more often than blue drivers (71\% vs. 64\%), while searches of red drivers recover contraband less often than searches of blue drivers (39\% vs. 44\%). 
Thus, the outcome and benchmark tests suggest that red drivers are being discriminated against in both (a) and (b). This is true in (a), because red drivers face a lower search threshold than blue drivers.
However, blue drivers are subject to the lower threshold in (b), contradicting the results of the benchmark and outcome tests.
     }}
     \label{fig:example_no_discrimination}
 \end{figure}
  
Fig.~\ref{fig:example_no_discrimination}a illustrates the setup described above for two hypothetical race groups, 
where the curves show race-specific \emph{signal distributions} (i.e., the distribution of guilt across all stopped
motorists of that race), and the 
vertical lines indicate race-specific search thresholds.
In this example, the red vertical line (at 30\%) is to the left of the blue vertical line (at 35\%), and so the red group, by definition, is being discriminated against.
Under our model, the search rate for each race equals the area under the group's signal distribution to the right of the corresponding race-specific threshold, 
which in this case is 71\% for the red group and 64\% for the blue group.
The hit rate (i.e., the search success rate) for each race equals the mean of the group's signal distribution conditional on being above the 
group's search threshold,
39\% for the red group and 44\% for the blue group.
The red group is thus searched at a higher rate (71\% vs.~64\%), and when searched, found to have contraband at a lower rate 
(39\% vs.~44\%) than the blue group.
Both the benchmark test (comparing search rates) and the outcome test (comparing hit rates) 
correctly indicate that the red group is being discriminated against.

\footnotetext{The depicted signal curves are beta distributions. The parameters for the red curves are: (a) $\alpha=10.2,\,\beta=18.8$; and (b) $\alpha=10.8,\,\beta=19.8$. The parameters for the blue curves are: (a) $\alpha=10.3,\,\beta=16.2$; and (b) $\alpha=2.1,\,\beta=4.1$.} 

\subsection{The problem of infra-marginality}

To illustrate the problem of infra-marginality, 
Fig.~\ref{fig:example_no_discrimination}b shows an alternative, hypothetical situation that is observationally equivalent to the one depicted in
Fig.~\ref{fig:example_no_discrimination}a,
meaning that the search and hit rates of the red and blue groups are exactly the same in both settings.
Accordingly, both the benchmark and outcome tests again suggest that the red group is being discriminated against.
In this case, however, blue drivers face a lower search threshold (25\%) than red drivers (30\%), and therefore the true discrimination present is exactly the opposite of the discrimination suggested by the outcome and benchmark tests.

What went wrong in this latter example?
It is easier in the blue group to distinguish between innocent and guilty individuals, 
as indicated by the signal distribution of the blue group having higher variance.
Consequently, those who are searched in the blue group are more likely to be guilty than those who are searched in the red group,
resulting in a higher hit rate for the blue group, throwing off the outcome test.
Similarly,
it is easier in the blue group to identify low-risk individuals, who need not be searched, in turn lowering the overall search rate of the group
and leading to spurious results from the benchmark test.
In this example, the search and hit rates are poor proxies for the search thresholds.

The key point about Fig.~\ref{fig:example_no_discrimination}b is that it is not a pathological case; 
to the contrary, it seems quite ordinary, and a variety of mechanisms could lead to this situation.
If innocent minorities anticipate being discriminated against, they might display the same behavior---nervousness and 
evasiveness---as guilty individuals, making it harder to distinguish those who are innocent from those who are guilty.
Alternatively, one group may simply be more experienced at concealing criminal activity, again making
it harder to distinguish guilty from innocent.
Given that one cannot rule out the possibility of such signal distributions arising in real-world examples 
(and indeed we later show that such cases do occur in practice), 
the benchmark and outcome tests are at best partial indicators of discrimination.
We address this so-called problem of infra-marginality by directly estimating the search thresholds themselves,
instead of simply considering the search and hit rates.

\subsection{Inferring search thresholds}

We now  describe our \emph{threshold test} for discrimination, which mitigates the problem of infra-marginality
in outcome tests.
For each stop $i$, we assume that we observe: 
(1) the race of the driver, $r_i$; 
(2) the department of the officer, $d_i$;
(3) whether the stop resulted in a search, indicated by $S_i \in \{0,1\}$;
and (4) whether the stop resulted in a ``hit'' (i.e., a successful search), indicated by $H_i \in \{0,1\}$.
Since a hit, by definition, can only occur if there was a search, $H_i \leq S_i$.
Given a fixed set of stops annotated with the driver's race and the officer's department,
we assume $S_i$ and $H_i$ are random outcomes resulting from a parametric process of search and discovery 
that formalizes the model of Section~\ref{sec:behavior}, described in detail below.
Our primary goal is to infer race-specific search thresholds for each department.
We interpret lower search thresholds for one group relative to another as evidence of discrimination.
For example, if we were to find black drivers face a lower search threshold than white drivers,
we would say blacks are being discriminated against.

We formalize this statistical problem in terms of a hierarchical Bayesian latent variable model. 
Our choice has two key benefits over natural alternatives.
First, in contrast to maximum likelihood estimation, Bayesian inference automatically yields robust estimates of uncertainty~\citep{gelman2014}, 
obviating the need for bootstrapping, which can be computationally expensive for complex models such as ours.
Second, hierarchical structure allows efficient pooling of evidence across departments. For example, if one race group is stopped only rarely in a given department, a hierarchical model can appropriately regularize department-level parameters toward state-level averages.

We next detail the generative model that underlies the threshold test. Consider a single stop of a motorist of race $r$ conducted by an officer in department $d$.
Upon stopping the driver, the officer assesses all the available evidence
and concludes the driver has probability $p$ of possessing contraband. 
Even though officers may make these judgements deterministically, 
there is uncertainty in who is pulled over in any given stop.
We thus model $p$ as a random draw from a race- and department-specific signal distribution,
which captures heterogeneity across stopped drivers.
This formulation sidesteps the omitted variables problem of benchmark tests 
by allowing us to express information from all unobserved covariates as variation in the signal distribution.
We can, in other words, think of the signal distribution as the marginal distribution over all unobserved variables.

We assume the signal $p$ is drawn from a beta distribution parameterized by its mean $\phi_{rd}$ (where $0 < \phi_{rd} < 1)$ and total 
count parameter $\lambda_{rd}$  (where $\lambda_{rd} > 0$).\footnote{In terms of the standard count parameters $\alpha$ and $\beta$ of the beta distribution,
$\phi = \alpha/(\alpha + \beta)$ and $\lambda = \alpha + \beta$.}
The $\phi_{rd}$ term is the overall probability that a stopped driver of race $r$ in department $d$ has contraband, while $\lambda_{rd}$ characterizes the heterogeneity across stopped drivers of that race in that department.
Turning to the search thresholds, we assume that officers in a department apply the same threshold $t_{rd}$ to all drivers of a given race,
but we allow these thresholds to vary by driver race and by department. Given the randomly drawn signal $p$, we assume officers deterministically decide to search a motorist if and only if $p$ exceeds $t_{rd}$;
and if a search is conducted, we assume that contraband is found with probability $p$.

As shown in Fig.~\ref{fig:example_no_discrimination}, different $(\phi_{rd},\lambda_{rd},t_{rd})$ tuples can result in the same observed search and hit rates. Thus, we require more structure in the parameters to ensure they are identified by the data.
To this end, we assume $\phi_{rd}$ and $\lambda_{rd}$ are functions of 
parameters that depend only on a motorist's race ($\phi_r$ and $\lambda_r$),
and those that depend only on an officer's department ($\phi_d$ and $\lambda_d$):

\begin{align}
    \phi_{rd} &= \textrm{logit}^{-1}(\phi_r + \phi_d) \label{eq:phi}
\end{align}
\noindent and
\begin{align}
    \lambda_{rd} &= \exp(\lambda_r + \lambda_d), \label{eq:lambda}
\end{align}

\noindent
where we set $\phi_d$ and $\lambda_d$ equal to zero for the largest department.\footnote{Without these constraints, 
the posterior distributions of the parameters would still be 
well-defined, but in that case the model would be identified by the priors rather than by the data. 
Moreover, without zeroing-out one pair of department parameters, 
the posterior distribution of $\phi_r$ would be highly correlated with that of $\phi_d$ (and likewise for $\lambda_r$ and $\lambda_d$),
which makes inference computationally difficult.}
As a result, if there are $D$ departments and $R$ races, the collection of $D \times R$ signal distributions is 
parameterized by $2(D+R-1)$ latent variables. 

In summary, for each stop $i$, the data-generating process for $(S_i, H_i)$ proceeds in three steps, as follows:
\begin{enumerate}
\item Given the race $r_i$ of the driver and the department $d_i$ of the officer, the officer
observes a signal $p_i \sim \text{beta}(\phi_{r_id_i}, \lambda_{r_id_i})$,
where $\phi_{r_id_i}$ and $\lambda_{r_id_i}$ are defined according to Eqs. \eqref{eq:phi} and \eqref{eq:lambda}.
\item $S_i = 1$ (i.e., a search is conducted) if and only if $p_i \geq t_{r_id_i}$.
\item If $S_i = 1$, then $H_i \sim \text{Bernoulli}(p_i)$; otherwise $H_i = 0$.
\end{enumerate}

\begin{figure}[t]
\centering
\begin{tikzpicture} 

\node[obs]                             (S_rd)           {$S_{rd}$} ;
\node[obs, right=of S_rd]      (H_rd)           {$H_{rd}$} ;
\node[latent, left=of S_rd, yshift=16mm]     (lambda_r)    {$\lambda_r$};
\node[latent, above=of lambda_r]     (phi_r)    {$\phi_r$};
\node[latent, above=of S_rd, xshift=9mm, yshift=8mm]    (t_rd)    {$t_{rd}$};
\node[latent, above=of t_rd, xshift=-9mm, yshift=8mm]     (mu_t_r)    {$\mu_{t_r}$};
\node[latent, above=of t_rd, xshift=9mm, yshift=8mm]     (sigma_t_r)    {$\sigma_{t_r}$};

\node[latent, right=of H_rd, yshift=16mm]     (lambda_d)    {$\lambda_d$};
\node[latent, above=of lambda_d]     (phi_d)    {$\phi_d$};
\node[latent, above=of phi_d]     (sigma_phi)    {$\sigma_{\phi}$};
\node[latent, right=of phi_d]     (mu_phi)    {$\mu_{\phi}$};
\node[latent, right=of lambda_d]     (mu_lambda)    {$\mu_{\lambda}$};
\node[latent, below=of mu_lambda]     (sigma_lambda)    {$\sigma_{\lambda}$};

\factoredge {} {lambda_r} {S_rd}
\factoredge {} {lambda_r} {H_rd}
\factoredge {} {phi_r} {S_rd}
\factoredge {} {phi_r} {H_rd}

\factoredge {} {t_rd} {S_rd}
\factoredge {} {t_rd} {H_rd}
\factoredge {} {mu_t_r} {t_rd}
\factoredge {} {sigma_t_r} {t_rd}

\factoredge {} {lambda_d} {S_rd}
\factoredge {} {lambda_d} {H_rd}
\factoredge {} {phi_d} {S_rd}
\factoredge {} {phi_d} {H_rd}

\factoredge {} {sigma_phi} {phi_d}
\factoredge {} {mu_phi} {phi_d}
\factoredge {} {mu_lambda} {lambda_d}
\factoredge {} {sigma_lambda} {lambda_d}
  
   \plate {plate1} {
    (S_rd)
    (H_rd)  
    (lambda_r)
    (phi_r) 
    (t_rd)  
    (mu_t_r)
    (sigma_t_r)
  } {$R$};
  
    \plate {plate2} {
    (S_rd)
    (H_rd)  
    (lambda_d)
    (phi_d) 
    (t_rd)
    (plate1.south)  
    } {$D$};
    
\end{tikzpicture}

\caption{\emph{Graphical representation of our generative model of traffic stops and searches. Observed search and hit rates are shaded, and unshaded nodes are latent variables that we infer from data.}} \label{fig:graphical_model}
\end{figure}
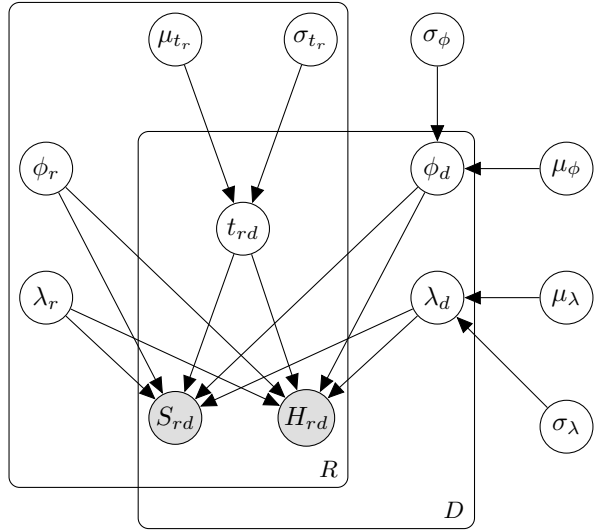

This generative process is parameterized by
 $\{\phi_r\}$, $\{\lambda_r\}$, $\{\phi_d\}$, $\{\lambda_d\}$ and $\{t_{rd}\}$.
To complete the Bayesian model specification, we put weakly informative
$\text{N}(0,2)$ priors on $\phi_r$ and $\lambda_r$,
and hierarchical priors on $\phi_d$, $\lambda_d$, and $t_{rd}$.
Specifically, we set 
$$\phi_d \sim \text{N}(\mu_{\phi}, \sigma_{\phi})$$
where $\mu_{\phi} \sim \text{N}(0,2)$ and $\sigma_{\phi} \sim \text{N}_+(0,2)$ (i.e., $\sigma_{\phi}$ has a half-normal distribution).
We similarly set 
$$\lambda_d \sim \text{N}(\mu_{\lambda}, \sigma_{\lambda})$$ 
where $\mu_{\lambda} \sim \text{N}(0,2)$ and $\sigma_{\lambda} \sim \text{N}_+(0,2)$.
Finally, for each race $r$, we put a logit-normal prior on every department's search threshold:
$$t_{rd} \sim \text{logit}^{-1}\left(\text{N}(\mu_{t_r}, \sigma_{t_r})\right)$$ 
where the race-specific hyperparameters $\mu_{t_r}$ and $\sigma_{t_r}$ have hyperpriors $\mu_{t_r} \sim \text{N}(0,2)$ and $\sigma_{t_r} \sim \text{N}_+(0,2)$.
This hierarchical structure allows us to make reasonable inferences even for departments with a relatively small number of stops.
We note that our results are robust to the exact specification of priors.\footnote{%
The results of our main analysis do not change if we use broader priors (e.g., $\text{N}(0,4)$), though broader priors come at the expense of longer inference times.
To see that our chosen prior structure is \emph{weakly} informative, consider the range of values within two standard deviations under each prior. For $\phi_r$, this encompasses a 2\% to 98\% chance of carrying contraband.
The search thresholds can likewise reasonably vary between 2\% and 98\%.
The prior on $\mu_\phi$ allows the average department to differ from the largest department by 4 points on the logit scale. In particular, if 20\% of people carry contraband in the largest department, the department mean can vary from 0.5\% to 93\%. The $\lambda$ parameters are exponentiated so they represent scaling factors: $\lambda_{rd}$ can reasonably be 50 times greater for one department than another.}
Figure~\ref{fig:graphical_model} shows this process represented as a graphical model~\citep{jordan2004}.

The number of observations $\mathcal{O} = \{(S_i, H_i)\}$ equals the number of stops---which could be in the millions---and 
so it can be computationally difficult to naively estimate the posterior distribution of the parameters.
We can, however, dramatically improve the speed of inference by re-expressing the model in terms 
of the total number of searches ($S_{rd}$) and hits ($H_{rd}$) for drivers of each race in each department:
\begin{align*}
S_{rd} & = \sum_{T_{rd}} S_i \\
H_{rd} & = \sum_{T_{rd}} H_i
\end{align*}
where $T_{rd} = \{i \, | \,  r_i = r \text{ and } d_i = d\}$.
Given that we fix the number of stops $n_{rd}$ of drivers of race $r$ in department $d$,
the quantities $\{S_{rd}\}$ and $\{H_{rd}\}$ are sufficient statistics for the process,
and there are now only $2DR$ quantities to consider, regardless of the number of stops.
This aggregation is akin to switching from Bernoulli to binomial response variables in a logistic regression model.

The distributions of $S_{rd}$ and $H_{rd}$ are readily computed for any parameter setting as follows.
Let $I_x(\phi, \lambda)$ be the cumulative distribution function for the beta distribution.
Then, 
\begin{equation*}
S_{rd} \sim \text{binomial}(p_{rd}, n_{rd})
\end{equation*}
where $p_{rd} = 1 - I_{t_{rd}}(\phi_{rd}, \lambda_{rd})$ is the probability that the signal is above the threshold.
Similarly, 
\begin{equation*}
H_{rd} \sim \text{binomial}(q_{rd}, S_{rd})
\end{equation*}
where for $p \sim \text{beta}(\phi_{rd}, \lambda_{rd})$, $q_{rd} = \mathbb{E}[p \:|\: p \geq t_{rd}]$ is the likelihood of finding contraband 
when a search is conducted.
A straightforward calculation shows that 
\begin{equation}
\label{eq:hit-rate}
    q_{rd} = \phi_{rd} \cdot \frac{1-I_{t_{rd}}\left(\mu_{rd},\, \lambda_{rd}+1\right)}{1-I_{t_{rd}}(\phi_{rd},\, \lambda_{rd})}
\end{equation}
where $\mu_{rd} = (\phi_{rd}\lambda_{rd}+1)/(\lambda_{rd}+1)$.
With this reformulation, it is computationally tractable to run the threshold test on large datasets.\footnote{%
A variant of the threshold test has recently been proposed by \citet{pierson2017} which can accelerate inference by
more than two orders of magnitude, allowing the test to scale to even larger datasets.
}

Having formally described our estimation strategy, we conclude by offering some additional intuition for our approach.
Each race-department pair has three key parameters: the threshold $t_{rd}$ and two parameters ($\phi_{rd}$ and $\lambda_{rd}$) that define the beta signal distribution.
Our model is thus in total governed by $3DR$ terms.
However, we only effectively observe $2 D R$ outcomes, the search and hit rates for each race-department pair.
We overcome this information deficit in two ways.
First, we restrict the form of the signal distributions according to Eqs.~\eqref{eq:phi} and \eqref{eq:lambda},
representing the collection of $DR$ signal distributions with $2(D+R-1)$ parameters.
With this restriction, the process is now fully specified by $2(D+R-1) + DR$ total terms, 
which is fewer than the $2 D R$ observations when $R \geq 3$ and $D \geq 5$.
Second, we regularize the parameters via hierarchical priors, which lets us efficiently pool information across races and
departments. 
In this way, we leverage heterogeneity across jurisdictions to simultaneously infer signal distributions and 
thresholds for all race-department pairs.

\section{An Empirical Analysis of North Carolina Traffic Stops}

Using the approach described above, we now test for discrimination in police searches of motorists stopped in North Carolina.

\subsection{The data}
We consider a comprehensive dataset of 9.5 million traffic stops conducted in North Carolina between January 2009 and December 2014 
that was obtained via a public records request filed with the state.
Several variables are recorded for each stop, including the race of the driver (white, black, Hispanic, Asian, Native American, or ``other''), 
the officer's department,
the reason for the stop, 
whether a search was conducted,
the type of search,
the legal basis for that search, 
and whether contraband (e.g., drugs, alcohol, or weapons) was discovered during the search.\footnote{In our analysis, 
``Hispanic'' includes anyone whose ethnicity was recorded as Hispanic, irrespective of their recorded race (e.g., it includes
both white and black Hispanics). 
} 
Due to lack of data, we exclude Native Americans from our analysis, who comprise fewer than 1\% of all stops; 
we also exclude the 1.2\% of stops where the driver's race was not recorded or was listed as ``other''. 

We say that a stop resulted in a search if any of four listed types of searches (driver, passenger, vehicle, or property) were conducted. 
There are five legal justifications for searches recorded in our dataset:
(1) the officer had probable cause that the driver possessed contraband;
(2) the officer had reasonable suspicion---a weaker standard than probable cause---that the driver presented a danger, 
and searched the passenger compartment of the vehicle to secure any weapons that may be present (a ``protective frisk'');
(3) the driver voluntarily consented to the officer's request to search the vehicle;
(4) the search was conducted after an arrest was made to look for evidence related to the alleged crime (a search ``incident to arrest'');
and (5) the officer was executing a search warrant.
There is debate over which searches should be considered when investigating discrimination. 
For example, \citet{engel2008} argue that because consent searches involve decisions by both officers and drivers, they should not be used to investigate possible discrimination in officer decisions; \citet{maclin2008} disagrees, claiming that consent searches give officers ``discretion to conduct an open-ended search with virtually no limits,'' and are thus an important way in which discrimination could occur.
In our primary analysis, we include all searches, 
regardless of the recorded legal justification.
We note, however, that our substantive results do not change if
we consider only probable cause searches, which all authors appear to include in their analysis;
our results also remain unchanged if 
we restrict to the set of probable cause, protective frisk, and consent searches, as \citet{hetey2016} suggest.

There are 287 police departments in our dataset, including city departments, departments on college campuses, 
sheriffs' offices, and the North Carolina State Patrol. 
We find that state patrol officers conduct 47\% of stops but carry out only 12\% of all searches, and recover only 6\% of all contraband found. 
State patrol officers search vastly less often than other officers, and the relatively few searches they do carry out are less successful.
Given these qualitative differences, we exclude state patrol searches from our primary analysis. 
We further restrict to the 100 largest local police departments (by number of recorded stops), 
which in aggregate comprise 91\% of all non-state-patrol stops.
We are left with 4.5 million stops that we use for our primary analysis.
Among this set of stops, 50\% of drivers are white, 40\% are black, 8.5\% are Hispanic, and 1.5\% are Asian.
The overall search rate is 4.1\%, and 29\% of searches turn up contraband.

\subsection{Results from benchmark and outcome tests}

\begin{table}[t]
\centering
\makebox[\textwidth][c]{
\begin{tabular}{lrrr}
  \toprule
Driver race & Stop count & Search rate & Hit rate \\ 
  \midrule
White & 2,227,214 & 3.1\% & 32\% \\ 
  Black & 1,810,608 & 5.4\% & 29\% \\ 
  Hispanic & 384,186 & 4.1\% & 19\% \\ 
  Asian & 67,508 & 1.7\% & 26\% \\ 
   \bottomrule
\end{tabular}}
\caption{\emph{Summary of the traffic stops conducted by the 100 largest police departments in North Carolina. Relative to white drivers, the benchmark test (comparing search rates) finds discrimination against blacks and Hispanics, while the outcome test (comparing hit rates) finds discrimination against blacks, Hispanics, and Asians.}} 
\label{tab:results}
\end{table}

We start with standard benchmark and outcome analyses of North Carolina traffic stops.
Table~\ref{tab:results} shows that the search rate for black drivers (5.4\%) and Hispanic drivers (4.1\%) 
is higher than for whites drivers (3.1\%).
Moreover, when searched, the rate of recovering contraband on 
blacks (29\%) and Hispanics (19\%) is lower than when searching whites (32\%).
Thus both the benchmark and outcome tests point to discrimination in search decisions against blacks and Hispanics. 
The evidence for discrimination against Asians is mixed. 
Asian drivers are searched less often than whites (1.7\% vs.~3.1\%), 
but these searches also recover contraband at a lower rate (26\% vs.~32\%). 
Therefore, relative to whites, the outcome test finds discrimination against Asians but the benchmark test does not. 

\begin{figure}[t!]
     \centering
         \includegraphics[width=\linewidth]{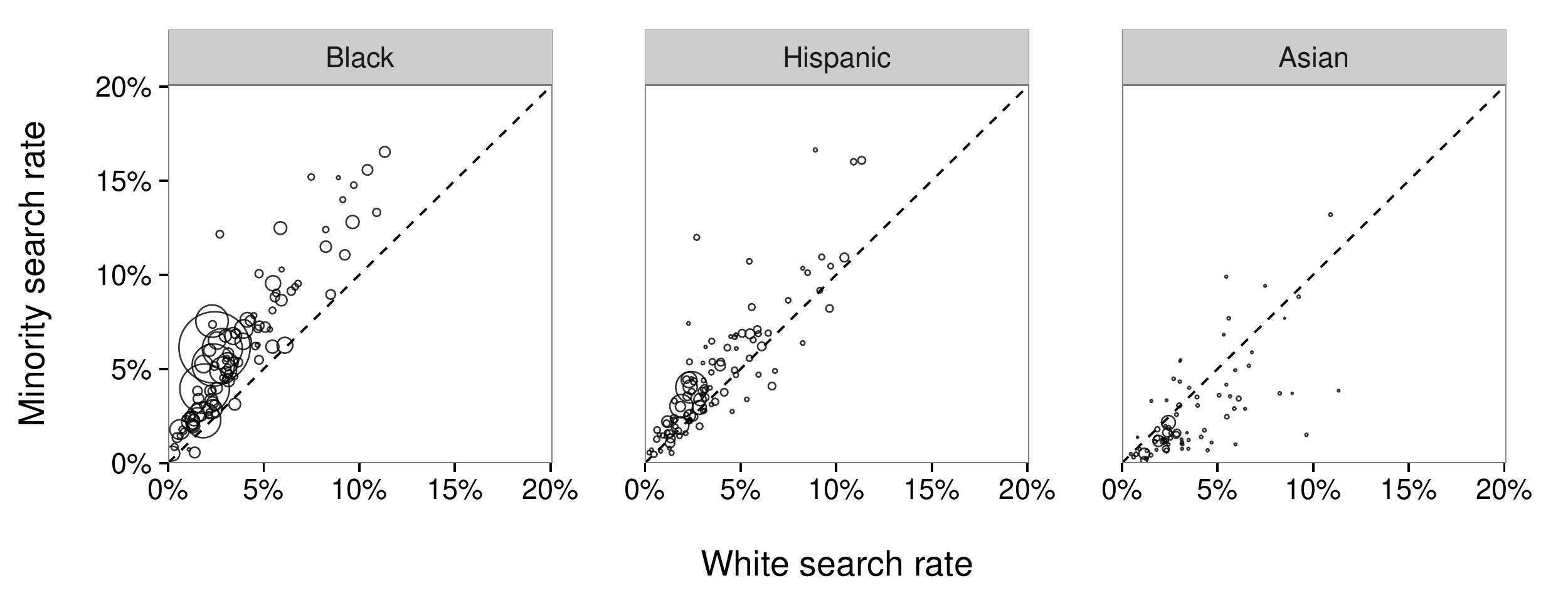} \\
         \vspace{5mm}
         \includegraphics[width=\textwidth]{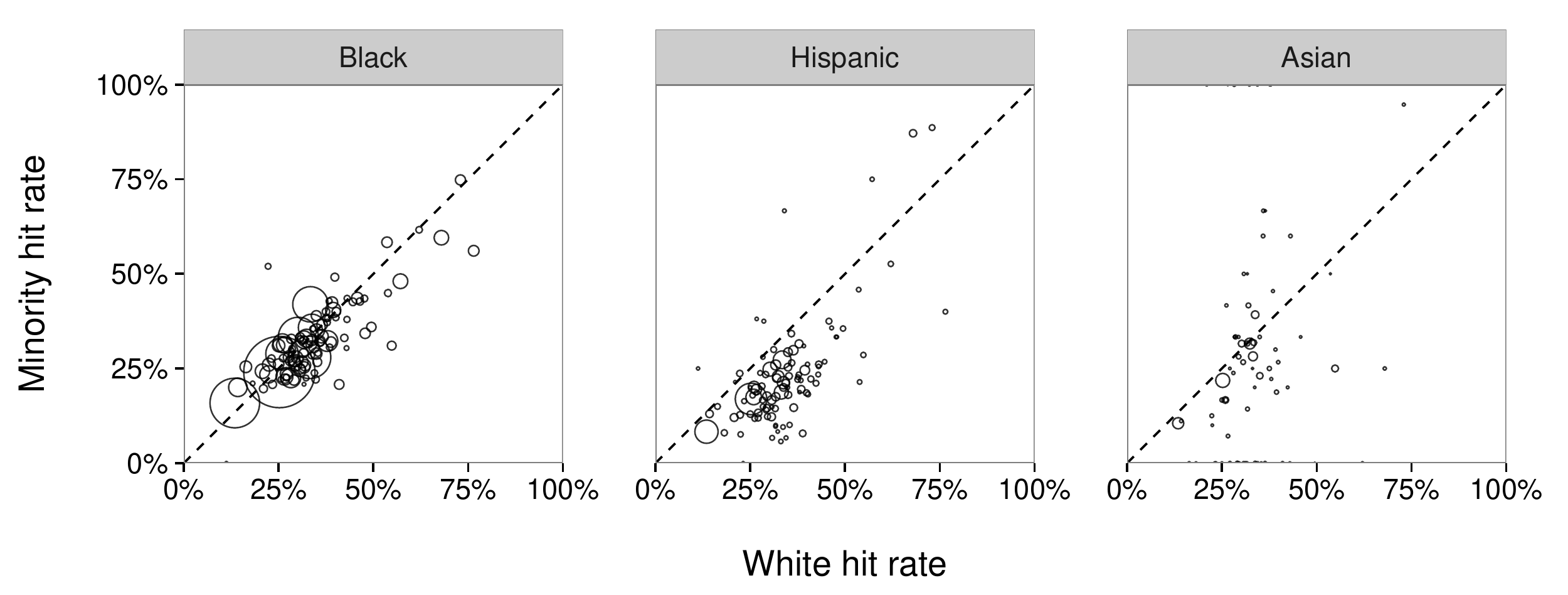}
     \caption{\emph{Results of benchmark and outcome tests on a department-by-department basis.
     Each point in the top panel compares search rates of minority and white drivers for a single department.
     In the vast majority of departments, blacks and Hispanics are searched at higher rates than whites.
     In the bottom panel, each point compares the corresponding department-level hit rates.
     While Hispanics have consistently lower hit rates than whites, black and white hit rates are comparable in many departments;
     the outcome test thus suggests an absence of discrimination against blacks in many departments.
     Points in all the plots are scaled to the number of times the minority race was stopped by the department.}}
     \label{fig:benchmark_outcome}
 \end{figure}

Adding resolution to these aggregate results, 
Fig.~\ref{fig:benchmark_outcome} compares search and hit rates for minorities and whites in each department.
In the vast majority of cases, the top panel shows that 
blacks and Hispanics are searched at higher rates than whites. 
Asians, however, are consistently searched at lower
rates than whites---indicating an absence of discrimination against Asians---in line with the aggregate results discussed above.
The department-level outcome analysis is shown in the bottom panel of Fig.~\ref{fig:benchmark_outcome}.
In most departments, when Hispanics are searched, they are found to have contraband less often than searched whites,
indicative of discrimination.
However, hit rates for blacks and Asians are comparable to, or even higher than, hit rates for whites in a substantial fraction of cases, 
suggesting a lack of discrimination against these groups in many departments.

Both the benchmark and outcome tests suggest discrimination against blacks and Hispanics in the majority of police departments, 
but also yield conflicting results in a significant number of cases.
For example, both tests are indicative of discrimination against blacks in 57 of the top 100 departments;
but in 42 departments, they offer ambiguous evidence, with one test pointing toward discrimination against black drivers while
the other indicates discrimination against white drivers. 
In one department both the outcome and benchmark tests point to discrimination against white drivers.
 
 \subsection{Results from the threshold test}

We next use our threshold test to infer race- and department-specific search thresholds.
Given the observed data, we estimate the posterior distribution of the search thresholds
via Hamiltonian Monte Carlo (HMC) sampling~\citep{neal1994,duane1987}, a form of Markov chain Monte Carlo sampling~\citep{metropolis1953}.
We specifically use the No-U-Turn sampler (NUTS)~\citep{hoffman2013} as implemented in Stan~\citep{carpenter2016},
an open-source modeling language for full Bayesian statistical inference.
To assess convergence of the algorithm, we sampled five Markov chains in parallel and computed the potential scale reduction factor $\hat{R}$~\citep{gelman1992}. 
We found that 2,500 warmup iterations and 2,500 sampling iterations per chain were sufficient for convergence,
as indicated by $\hat{R}$ values less than $1.05$ for all parameters, as well as by visual inspection of the trace plots.

\begin{figure}[t]
	\centering
 	\includegraphics[width=\textwidth]{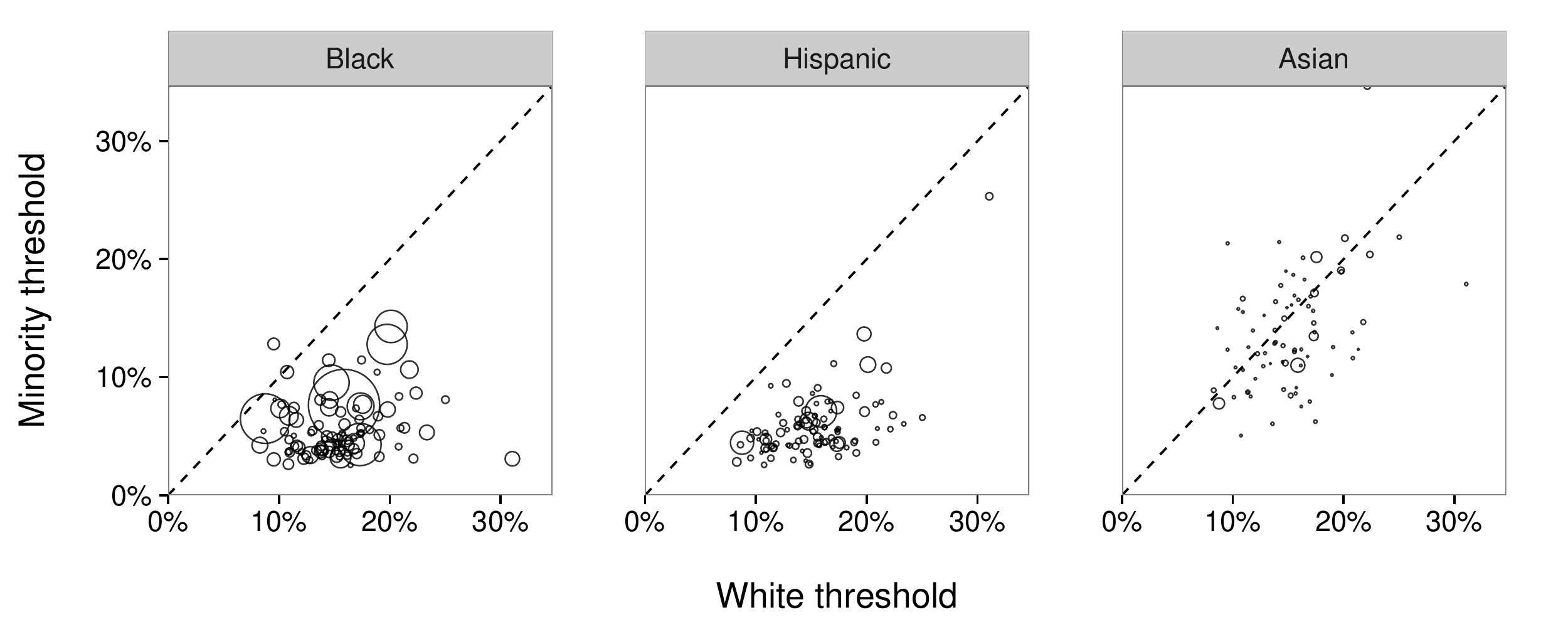}
	\caption{\emph{Inferred search thresholds in the 100 largest North Carolina police departments. 
     Each point compares the search thresholds applied to minority and white drivers in a department,
     where points are scaled to the number of times the minority race was stopped by the department.
     In nearly every department, black and Hispanic drivers face lower search thresholds than whites, suggestive of discrimination.}}
      \label{fig:thresholds}
 \end{figure}
 
 \begin{figure}[t]
	 \centering
         \includegraphics[width=.48\linewidth]{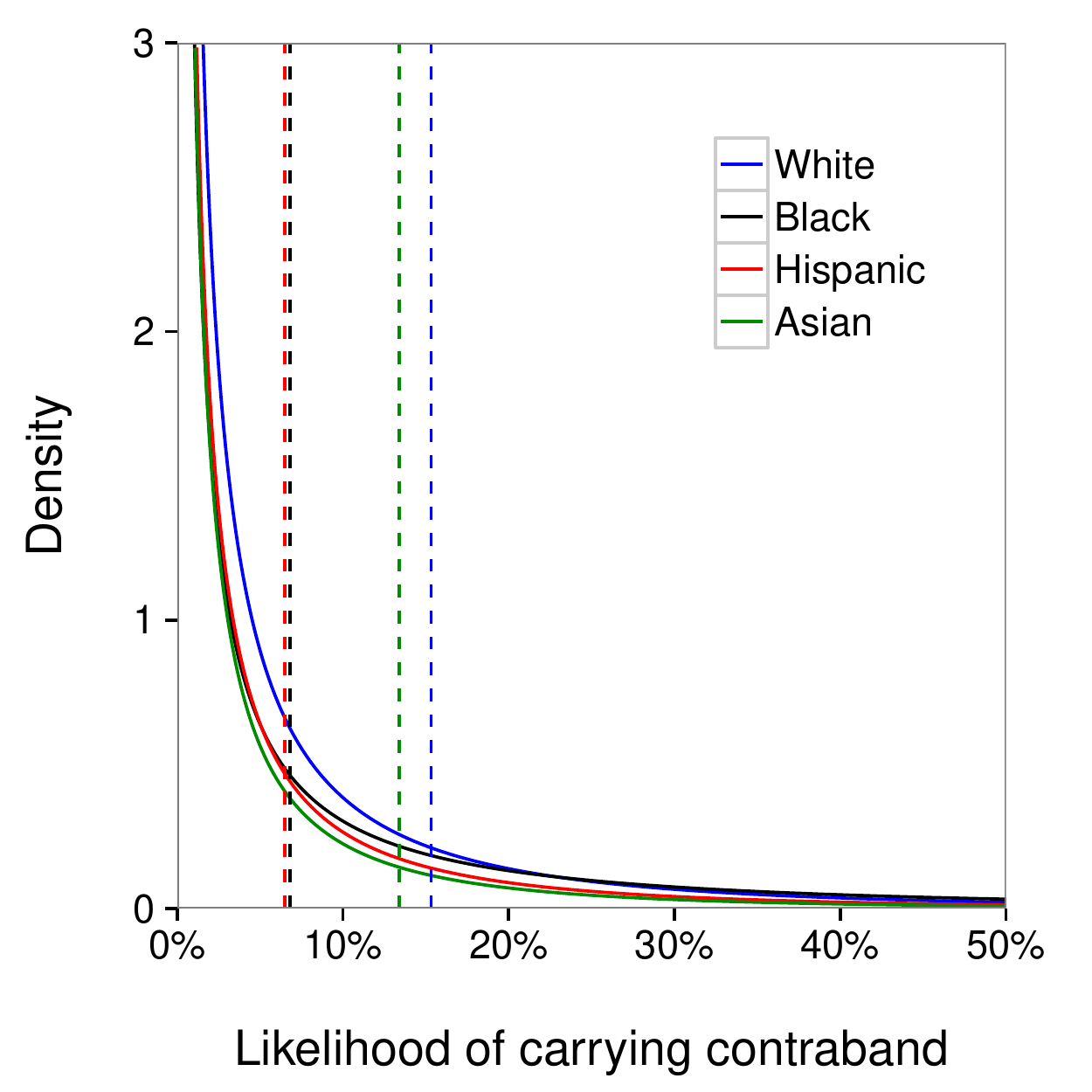}
          \caption{\emph{Race-specific search thresholds and signal distributions, averaged over all departments and where
     we weight by the total number of stops conducted by the department.
      We find that black and Hispanic drivers face substantially lower search thresholds than white and Asian drivers.}}
      \label{fig:signal}
 \end{figure}

Figure~\ref{fig:thresholds} shows the posterior mean search thresholds for each race and department.
Each point in the plot corresponds to a department, and compares the search threshold for whites (on the $x$-axis) to
that for minorities (on the $y$-axis).
In nearly all the departments we consider,
the inferred search thresholds for black and Hispanic drivers are lower than for whites, 
suggestive of discrimination against these groups.
For Asians, in contrast, the inferred search thresholds are generally in line with those of whites, 
indicating an absence of discrimination against Asians in search decisions.

Figure \ref{fig:signal} displays the average, state-wide inferred signal distributions and thresholds for whites, blacks, Hispanics, and Asians.
These averages are computed by weighting the department-level results by the number of stops in the department.
Specifically, the overall race-specific threshold $t_r $ is given by $\left(\sum_d t_{rd} \cdot n_{d} \right)/ \sum_d n_{d}$,
where $n_d$ is the number of stops in department $d$.
Similarly, the aggregate signal distributions show the department-weighted distribution of probabilities of possessing contraband.
As is visually apparent, and also summarized in Table~\ref{tab:threshold_results}, the inferred thresholds for searching whites (15\%) and Asians (13\%) 
are significantly higher than the inferred thresholds for searching blacks (7\%) and Hispanics (6\%).
These thresholds are estimated to about $\pm 2\%$, 
as indicated by the 95\% credible intervals listed in Table~\ref{tab:threshold_results}. 

\begin{table}[t]
\centering
\makebox[\textwidth][c]{
\begin{tabular}{lrr}
  \toprule
Driver race & Search threshold & 95\% credible interval \\ 
  \midrule
White & 15\% & (14\%, 16\%) \\ 
  Black & 7\% & (3\%, 10\%) \\ 
  Hispanic & 6\% & (5\%, 8\%) \\ 
  Asian & 13\% & (11\%, 16\%) \\ 
   \bottomrule
\end{tabular}}
\caption{\emph{Inferred search thresholds for stops conducted by the 100 largest police departments in North Carolina. For each race group, we report the average threshold across departments, weighting by the number of stops conducted by the department. We find black and Hispanic drivers face lower search thresholds than white and Asian drivers.}} 
\label{tab:threshold_results}
\end{table}

\begin{figure}[t]
     \centering
     \includegraphics[width=0.48\textwidth]{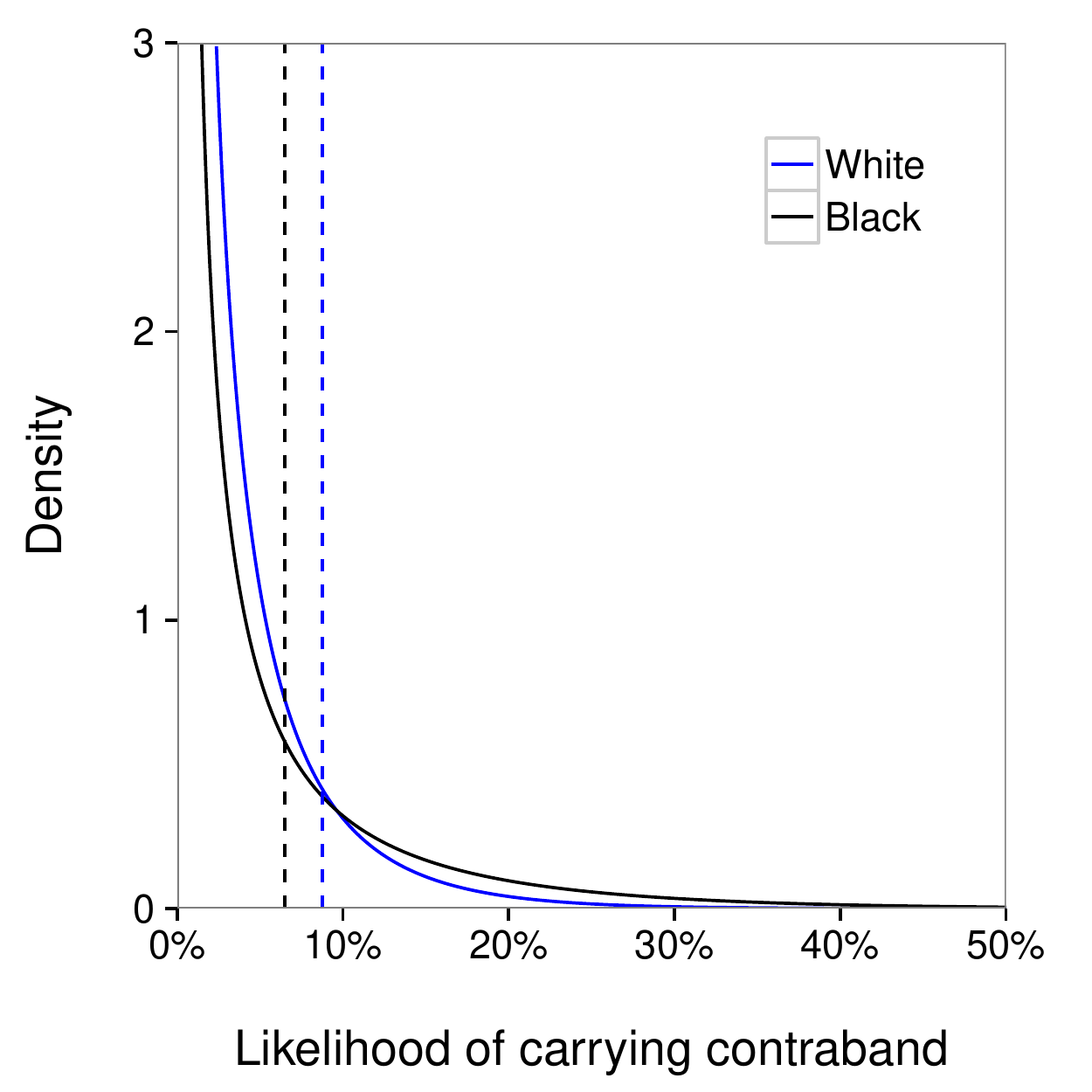}
     \captionof{figure}{\emph{Inferred search thresholds and signal distributions for black and white drivers stopped by the Raleigh Police Department, illustrating the problem of infra-marginality. The heavier tail of the black signal distribution means that searches of blacks have a higher hit rate despite black drivers facing a lower search threshold than whites. Hence, the outcome test concludes white drivers are being discriminated against, whereas the threshold test finds discrimination against black drivers.}}
     \label{fig:raleigh}
 \end{figure}

\subsection{The effects of infra-marginality}
Why is it that the threshold test shows consistent discrimination against blacks and Hispanics when
benchmark and outcome analysis suggest a more ambiguous story?
To understand this dissonance, we examine the specific case of the Raleigh Police Department, the second largest 
department in North Carolina by number of stops recorded in our dataset. 
Black drivers in Raleigh are searched at a higher rate than whites (4\% vs. 2\%),
but when searched, blacks are also found to have contraband at a higher rate (16\% vs. 13\%).
The benchmark and outcome tests thus yield conflicting assessments of whether black drivers face discrimination.
Figure \ref{fig:raleigh} shows the inferred signal distributions and thresholds for white and black drivers in Raleigh, 
and sheds light on these seemingly contradictory results. 
The signal distribution for black drivers has a heavier right tail---for example, there is four times more mass above 20\% than in the white distribution. 
This suggests that officers can more easily determine which black drivers are carrying contraband, 
which causes their searches of blacks to be more successful than their searches of whites. 
In spite of the higher hit rate for black drivers, 
we find that blacks still face a lower search threshold (6\%) than whites (9\%), suggesting discrimination against blacks. 

Despite the theoretical advantages of the threshold test, it is difficult to know for sure whether 
the threshold test or the outcome test better reflects decision making in Raleigh.
We note, though, three reasons that suggest the threshold test is the more accurate one.
First, looking at Hispanic drivers in Raleigh, both the benchmark and outcome tests indicate they face discrimination.
Hispanic drivers are searched more often than whites (3\% vs. 2\%), and are found to have contraband less often (11\% vs. 13\%).
The threshold test likewise finds evidence of discrimination against Hispanics.
The outcome test applied to black drivers is thus the odd one out:
the benchmark, outcome, and threshold tests all point to discrimination against Hispanic drivers, 
and the benchmark and threshold tests suggest discrimination against black drivers.
Second, the outcome test indicates not only an absence of discrimination, but that white drivers face substantial bias;
while possible, that conclusion is at odds with past empirical research on traffic stops~\citep{epp2014}. 
Finally, the data suggest a compelling explanation for the heavier tail in the inferred signal distribution for black drivers: 
stopped blacks may be
more likely than whites to carry contraband in plain
view, as indicated by the fact that 
stops of blacks are three times more likely to end in searches based on ``observation of suspected contraband''.\footnote{%
Searches based on ``observation of suspected contraband'' yield contraband in 52\% of cases,
which is substantially higher than searches premised on other factors.
}
The Raleigh Police Department thus appears to be a real-world example in which infra-marginality leads the outcome test to produce spurious results.

\subsection{Model checks}
\label{sec:validation}

We now evaluate in more detail how well our analytic approach explains the observed patterns in the North Carolina traffic stop data,
and examine the robustness of our conclusions to violations of the model assumptions.

\paragraph{Posterior predictive checks.}

We begin by investigating the extent to which the fitted model yields race- and department-specific search and hit rates that are in line with the observed data.
Specifically, for each department and race group, we compare the observed search and hit rates to their expected values under the assumed data-generating process 
with parameters drawn from the inferred posterior distribution. Such \emph{posterior predictive checks}~\citep{gelman2014,gelman1996} 
are a common approach for identifying and measuring systematic differences between a fitted Bayesian model and the data.

We compute the posterior predictive search and hit rates as follows. 
During model inference, our Markov chain Monte Carlo sampling procedure yields 2,500 draws from the
joint posterior distribution of the parameters. 
For each parameter draw---consisting of $\{\phi_r^*\}$, 
$\{\lambda_r^*\}$, $\{\phi_d^*\}$, $\{\lambda_d^*\}$ and $\{t_{rd}^*\}$---we analytically compute 
the search and hit rates $s_{rd}^*$ and $h_{rd}^*$ for each race-department pair implied by the data-generating process
with those parameters.
Finally, we average these search and hit rates over all 2,500 posterior draws.

\begin{figure}[t]
\makebox[\textwidth][c]{
    \begin{subfigure}[t]{.55\linewidth}
		\centering
		\includegraphics[width=\textwidth]{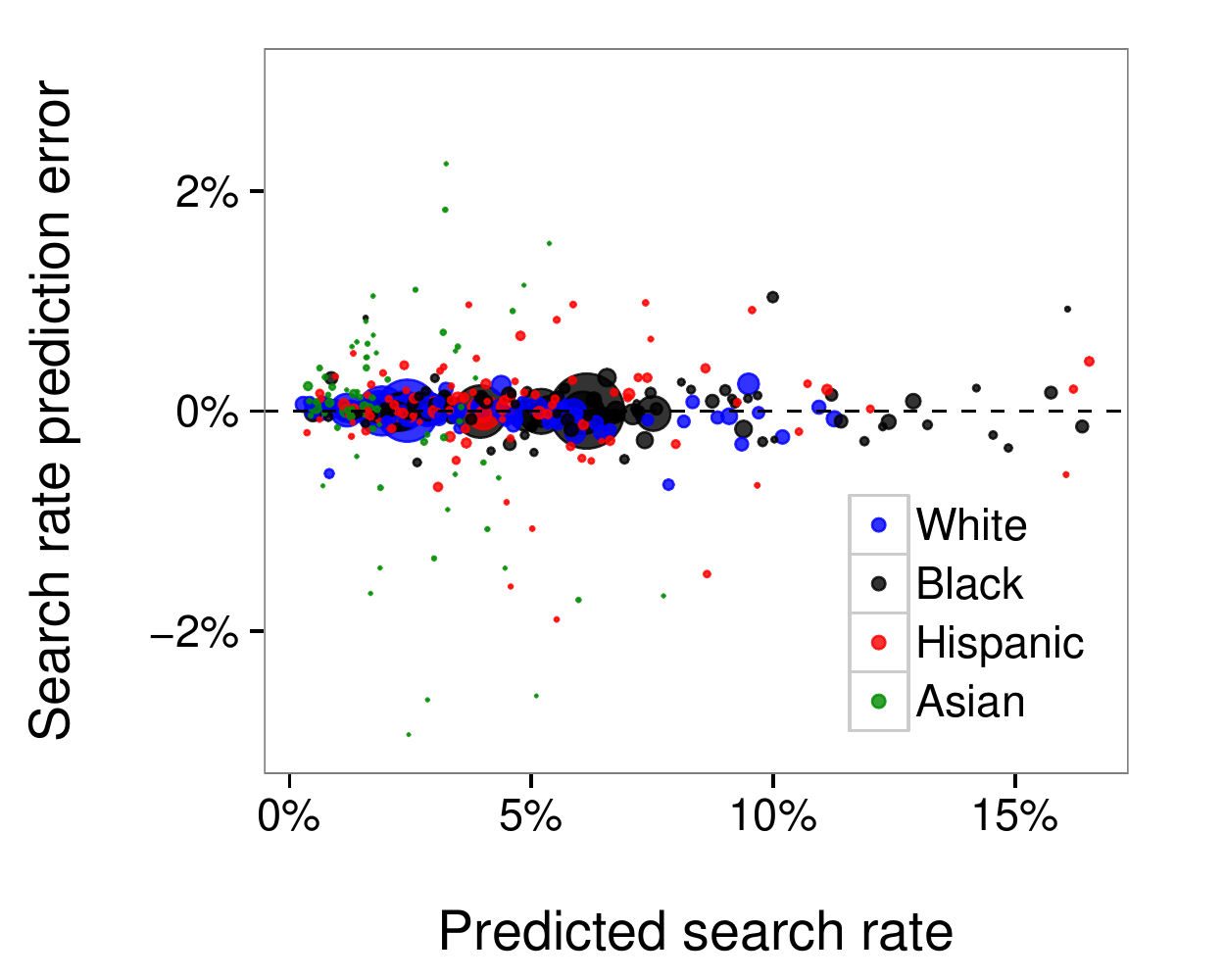}
		\caption{}	
		\label{fig:search_rate_ppc}
    \end{subfigure}
    \begin{subfigure}[t]{.55\linewidth}
        \centering
        \includegraphics[width=\linewidth]{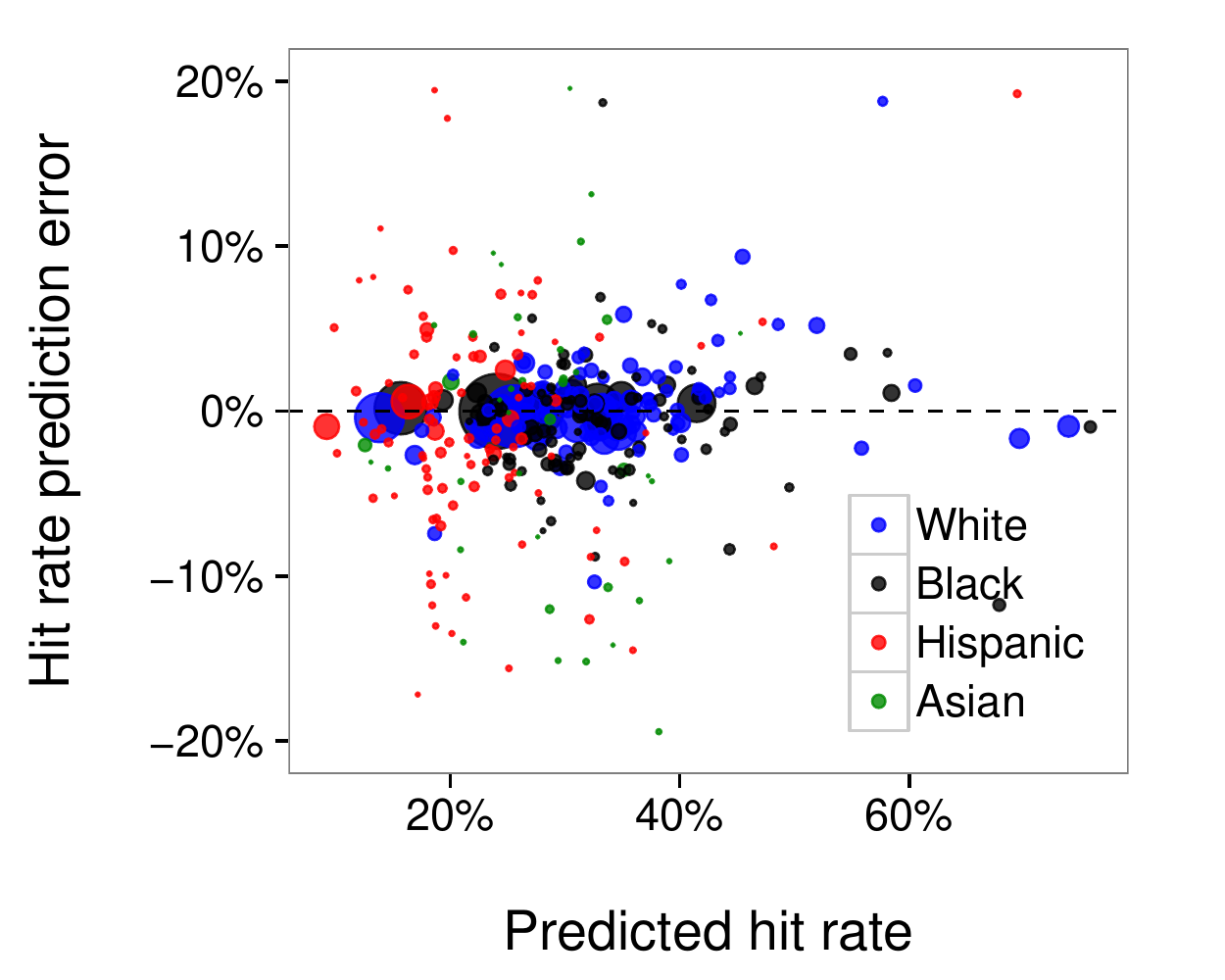}
        \caption{}
        \label{fig:hit_rate_ppc}
    \end{subfigure}}
    \caption{\emph{Comparison of model-implied search and hit rates to the actual, observed values. Each point is a race-department pair, with
    points sized by number of stops. The plots show that the fitted model captures key features of the observed data. The root mean squared prediction error (weighted by stop count) is 0.1\% for search rate and is 2.9\% for hit rate.}}
    \label{fig:ppc}
\end{figure}

Figure~\ref{fig:ppc} compares the model-predicted search and hit rates to the actual, observed values. 
Each point in the plot corresponds to a single race-department group, where groups are sized by number of stops.
The fitted model recovers the observed search rates almost perfectly across races and departments.
The fitted hit rates also agree with the data quite well,
with the largest groups exhibiting almost no error.
These posterior predictive checks thus indicate that the fitted model captures key features of the observed data.

\paragraph{Heterogeneous search thresholds.}

\begin{figure}[t]
     \centering
     \includegraphics[height=3in]{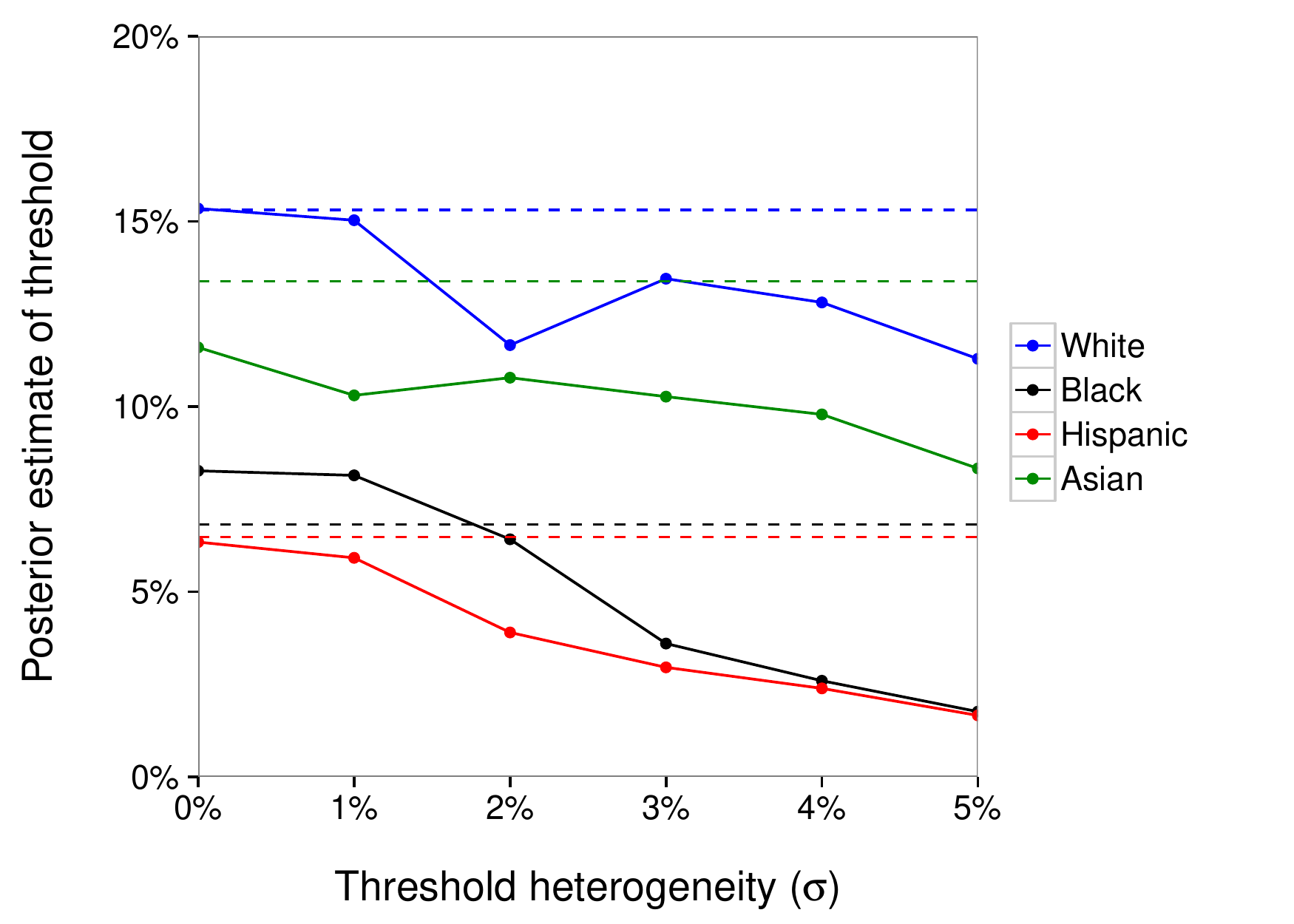}
     \captionof{figure}{\emph{Inferred race-specific search thresholds for synthetic data generated under a model in which thresholds randomly 
     vary from one stop to the next. The dashed horizontal lines show the average of the thresholds used to generate the data.
     Model inferences are largely robust to stop-level heterogeneity in search thresholds.
     }}
     \label{fig:noise}
 \end{figure}

Our behavioral model assumes that there is a single search threshold for each race-department pair.
In reality, officers within a department might apply different thresholds,
and even the same officer might vary the threshold he or she applies from one stop to the next.
Moreover, officers only observe noisy approximations of a driver's likelihood of carrying contraband;
such errors can be equivalently recast as variation in the search threshold applied to the true probability.

To investigate the robustness of our approach and results to such heterogeneity, 
we examine the stability of our inferences on synthetic datasets derived from a generative process with varying thresholds.
Specifically, we start with the model fit to the actual data and then proceed in four steps.
First, for each observed stop, we draw a signal $p$ from the inferred signal distribution for the department $d$ in which the stop occurred and
the race $r$ of the motorist.
Second, we set the stop-specific threshold to $T \sim \text{N}(t_{rd}, \sigma)$, where 
$t_{rd}$ is the inferred threshold, and $\sigma$ is a parameter we set to control the degree of heterogeneity in the thresholds.
Third, we assume a search occurs if and only if $p \geq T$, and if a search is conducted, we assume contraband is found with probability $p$.
Finally, we use our modeling framework to infer new search thresholds $t_{rd}'$ for the synthetic dataset.
Figure~\ref{fig:noise} plots the result of this exercise for $\sigma$ varying between 0 and 0.05. It shows that the inferences 
are relatively stable throughout this range, and in particular, that there is a persistent gap between whites and Asians compared to blacks and Hispanics.
We note that a five percentage point change in the thresholds is quite large. 
For example, decreasing the search threshold of blacks by five points in each department would more than triple the overall state-wide search rate of blacks.

\paragraph{Omitted variable bias.}
As we discussed in Section~\ref{sec:model}, our approach is robust to unobserved heterogeneity that affects the signal,
since we effectively marginalize over any omitted variables when estimating the signal distribution.
However, we must still worry about systematic variation in the thresholds that is correlated with race.
For example, if officers apply a lower search threshold at night, and black drivers
are disproportionately likely to be stopped at night, then blacks would, on average, experience a lower search threshold
than whites even in the absence of discrimination.
Fortunately, as a matter of policy, only a limited number of factors may legitimately affect the search thresholds, 
and many---but not all---of these are recorded in the data.
As a point of comparison, there are a multitude of hard-to-quantify factors (such as socio-economic indicators, or behavioral cues) that may, 
and likely do, affect the signal, but these should not affect the threshold.

\begin{figure}[t!]
     \centering
        \includegraphics[height=2.5in]{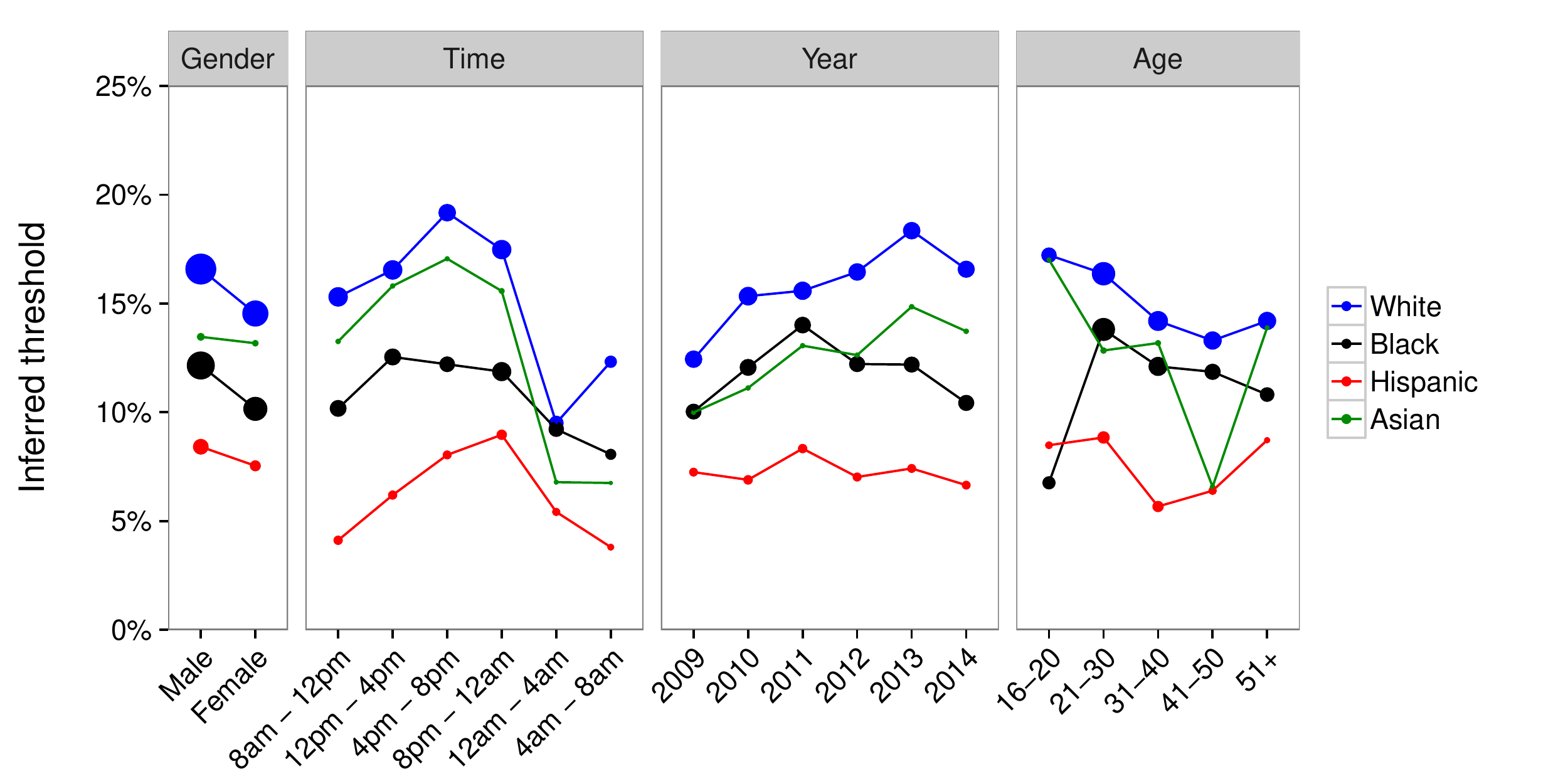} 
     \captionof{figure}{\emph{
	 Inferred search thresholds by race when the model is fit separately on various subsets of the data.
     Points indicate posterior means and are sized according to the number of stops in the subset.
     We consistently observe that blacks and Hispanics face lower search thresholds than whites. }}
     \label{fig:disaggregation}
 \end{figure}
 
Our model already explicitly accounts for search thresholds that vary by department.
We now examine the robustness of our results when adjusting for possible variation across year, time-of-day, age, and gender of the driver.\footnote{Gender, like race, is generally not considered a valid criterion for altering the search threshold, though for completeness we still examine its effects on our conclusions.}
Specifically, we disaggregate our primary dataset by year (and, separately, by time-of-day, by age, and by gender), and then independently run the threshold test on each component.\footnote{4.7\% of stops are recorded as occurring exactly at midnight, whereas 0.2\% are listed at 11pm and 0.1\% at 1am. 
It is likely that nearly all of the midnight stops are recorded incorrectly, and so we exclude these from our time-of-day analysis.
Similarly, in a small fraction of stops (0.1\%) the driver's age is recorded as either less than 16 or over 105 years old; we exclude these from our age analysis.}
Figure~\ref{fig:disaggregation} shows the results of this analysis, and illustrates two points.
First, we find that the inferred thresholds do indeed vary across the different subsets of the data.
Second, in every case, the thresholds for searching blacks and Hispanics are lower
than the threshold for searching whites, corroborating our main results.

In addition to the factors considered above, 
officers may legally apply a lower search threshold 
in situations involving officer safety.
In particular, ``protective frisks'' require only
 \emph{reasonable suspicion}, a lower standard of evidence than the \emph{probable cause} requirement that applies to most searches.
Further, probationers in North Carolina are subject to the reasonable suspicion standard regardless of safety issues.
Similarly, searches ``incident to arrest'' are often carried out 
as a matter of policy before transporting arrestees, and so such searches
may have a near-zero threshold.
If stopped black and Hispanic drivers are 
more likely than whites to fall into these categories
(e.g., if blacks and Hispanics are more likely to be on probation),
then the lower average search thresholds we find for minorities 
may not be the product of discrimination.
To test for this possibility, we now say a ``search'' has occurred only if:
(1) the basis for the search is recorded as ``probable cause''; 
and (2) ``other official info'' was not indicated as a precipitating factor.
The latter restriction is intended to exclude searches triggered by a driver's probation status.\footnote{%
The five search categories recorded in our data are:
``probable cause'', ``protective frisk'', ``consent'', ``incident to arrest'', 
and ``warrant''.
Searches of probationers in North Carolina require only reasonable suspicion---not probable cause---but that classification in not among the listed options, and so officers might still mark ``probable cause'' in these situations.
We infer whether a search was predicated on probation status by examining the factors listed as triggering the action, which may be any combination of: ``erratic/suspicious behavior'', ``observation of suspected contraband'', ``suspicious movement'', ``informant tip'', ``witness observation'', and ``other official info''.
The North Carolina Department of Public Safety
was unable to clarify the meanings of these options,
but it seems plausible that officers would mark ``other official info'' 
when a search is triggered by a driver's probation status.
Our results are qualitatively unchanged regardless of whether we include or exclude ``other official info'' searches.
}
Repeating our analysis with searches redefined in this way,
we find the basic pattern still holds:
blacks and Hispanics are searched at lower thresholds (8\% and 21\%, respectively) than whites and Asians (40\% and 38\%, respectively).
The inferred thresholds are higher than in our primary analysis---as expected, since 
we restricted to searches subject to a higher standard---but the gap remains.

A final potential confound is that search thresholds may vary by the severity of the contraband an officer believes could be present.
For example, if officers have a lower threshold for searching drivers when they suspect possession of cocaine rather than marijuana, and
black and Hispanic drivers are disproportionately likely to be suspected of carrying cocaine, then the threshold test could  
mistakenly infer discrimination where there is none.
Unfortunately, the suspected offense motivating a search is not recorded in our data, and so we cannot directly test for such an effect,
constituting one important limit of our statistical analysis.

\paragraph{Placebo tests.}

\begin{figure}[t!]
     \centering
     \includegraphics[height=3in]{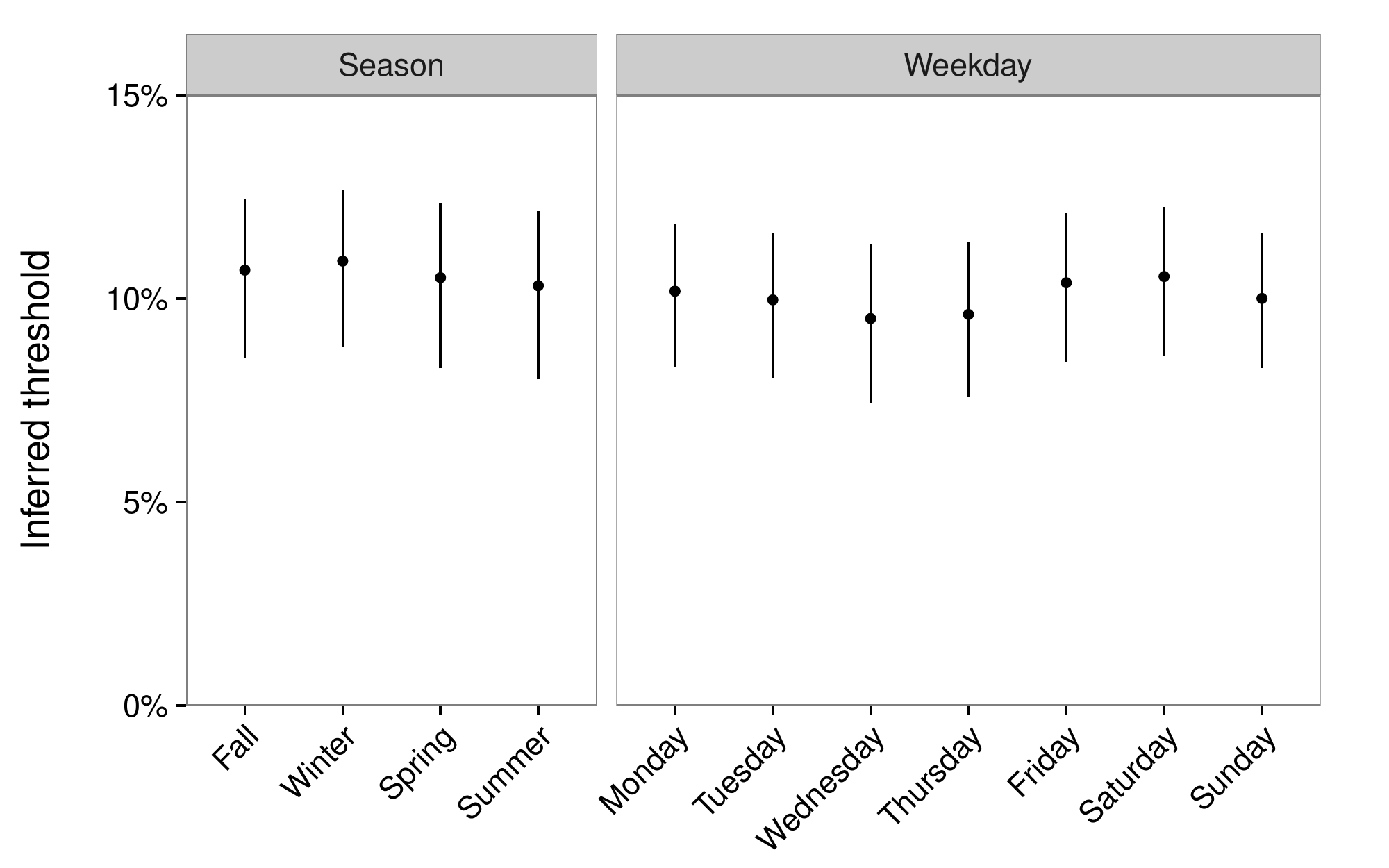}
     \captionof{figure}{\emph{Results of placebo tests, in which we examine how search thresholds vary by season and day-of-week.
     Points show the posterior means, and the bars indicate 95\% credible intervals.
     The threshold test accurately suggests a lack of ``discrimination'' in these cases. 
     }}
     \label{fig:placebo}
 \end{figure}

Finally, we conduct two \emph{placebo tests}, where we rerun our threshold test with race replaced by day-of-week, and separately, 
with race replaced by season.
The hope is that the threshold test accurately captures a lack of ``discrimination'' based on these factors.
Figure~\ref{fig:placebo} shows that the model indeed finds that the threshold for searching individuals is 
relatively stable by day-of-week, with largely overlapping credible intervals.
We similarly find only small differences in the inferred seasonal thresholds.
We note that some variation is expected, as officers might legitimately apply slightly different search standards
throughout the week or year.

																																																																															\section{Conclusion}

Theoretical limitations with the two most widely used tests for discrimination---the benchmark and outcome tests---have 
hindered investigations of bias.
Addressing this challenge, we have developed a new statistical approach to detecting discrimination
that builds on the strengths of the benchmark and outcome tests and that mitigates the shortcomings of both.
On a dataset of 4.5 million motor vehicle stops in North Carolina, our threshold test
suggests that black and Hispanic motorists face discrimination in search decisions.
Further, by specifically examining the Raleigh Police Department, we find that the problem of infra-marginality
appears to be more than a theoretical possibility, and may have caused the outcome test to mistakenly conclude 
that officers discriminated against white drivers. 

Our empirical results appear robust to reasonable violations of the model assumptions, 
including noise in estimates of the likelihood a driver is carrying contraband.
We have also attempted to rule out some of the more obvious legitimate reasons for which thresholds might vary, 
including search policies that differ across department, year, or time of day.
However, as with all tests of discrimination, there is a limit to what one can conclude from 
such statistical analysis alone. 
For example, if search policies differ not only across but also within department, 
then the threshold test could mistakenly indicate discrimination where there is none. 
Such within-department variation might result from explicit policy choices, or as a by-product of 
deployment patterns;
in particular,
the marginal cost of conducting a search may be lower in heavily policed neighborhoods, 
potentially justifying a lower search threshold in those areas.
Additionally, if officers suspect more serious criminal activity when searching
black and Hispanic drivers compared to whites, then the lower inferred search thresholds for these groups may be the result of non-discriminatory factors.
To a large extent, such limitations apply equally to past tests of discrimination,
and as with those tests, caution is warranted when interpreting the results.

Aside from police practices, the threshold test could be applied to study discrimination in a variety of settings
where benchmark and outcome analysis is the status quo, including lending, hiring, and 
publication decisions.
Looking forward, we hope our methodological approach spurs further investigation 
into the theoretical properties of statistical tests of discrimination, as well as their practical application.

\bibliographystyle{plainnat}
\bibliography{references}

\end{document}